\documentclass[twocolumn,showpacs,preprintnumbers,amsmath,amssymb]{revtex4} 

\usepackage{graphicx}
\usepackage{dcolumn}
\usepackage{bm}

\newcommand{\pabl}[2]{\frac{\partial #1}{\partial #2}}

\renewcommand{\d}{{\rm d}} 
\begin{document} 
 
\preprint{accepted by Physical Review B, 2002}
 
\title{Dynamical bistability in quantum dot structures: 
The role of Auger processes} 
 
\author{A. Rack, R. Wetzler, A. Wacker\email{wacker@physik.tu-berlin.de}, 
and E. Sch{\"o}ll} 
 
\affiliation{Institut f{\"u}r Theoretische Physik, 
Technische Universit{\"a}t Berlin, Hardenbergstr.~36, 10623~Berlin, Germany} 
\date{\today} 
 
\begin{abstract} 
Bistability in quantum dot structures is examined by a 
drift-diffusion model in combination with electron capture and emission 
processes. Our simulations provide  
a dynamic scenario with extremely long switching times 
of the order of months and  the results are in good agreement with the 
experimental findings of Yusa and Sakaki  
[Appl.\ Phys.\ Lett. {\bf 70}, 345 (1997)]. 
The analysis of the data supports the 
importance of Auger capture processes for quantum dots. 
\end{abstract} 
 
\maketitle 
 
\section{Introduction} 
Semiconductor quantum dots (QDs) formed by self-organized 
growth processes \cite{BIM99} 
allow for the spatial confinement of electrons 
on a scale of 10 nm. This suggests that these structures 
may serve as extremely small memory devices.
A first step in this direction is the investigation 
of specially grown structures where the charging of a 
layer of QDs could be detected via its influence on the 
conductivity of a two dimensional electron gas \cite{YUS97,FIN98}. 
These experiments showed a pronounced bistability (up to 
room temperature \cite{KOI00}) depending on the 
history of the bias sweep. The sensitivity to light illumination 
suggests a variety of applications in photoelectronic devices 
\cite{BOE99,SHI99}. 
 
A key issue towards 
a detailed understanding of these experiments is the question 
if the observed bistability is of transient nature,  
and the order of the time scales involved. 
In order to elucidate this point we performed a 
simulation of the experiment from Ref.~\onlinecite{YUS97} applying 
a drift-diffusion model combined 
with the generation-recombination (GR) kinetics of the QDs. 
Our results show that the experiments can be well described 
if Auger processes constitute the dominant GR-processes. 
Therefore this type of experiments may serve as an additional tool 
to shed light into the controversial issue on the nature of electron  
capture in quantum dots \cite{FEL01,RAY00,HEI97c}.  
 
\section{The Model} 

\begin{figure}
\includegraphics[width=0.8\columnwidth]{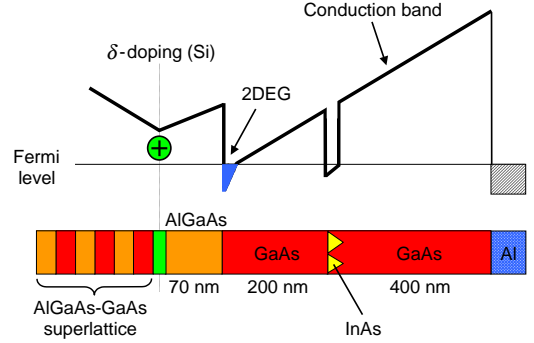} 
\caption[a]{Sketch of the sample used in Ref.~\onlinecite{YUS97} 
  together with the resulting conduction band profile at zero bias.} 
\label{Figsample} 
\end{figure}

Here we consider the structure used in the experiment of 
Ref.~\onlinecite{YUS97} as sketched in Fig.~\ref{Figsample}. 
From the experimentally determined size of the QDs 
(height $h=5\ \mathrm{nm}$ and diameter $20\ \mathrm{nm}$) 
we estimate a ground state binding energy of  
$E_{\mathrm{b}}= 250\ \mathrm{meV}$ \cite{STI99}. 
We use a QD density $N_{\mathrm{QD}}= 7.5 \times 10^{10}/\mathrm{cm}^2$ .

We model the transport in the conduction band by a drift-diffusion 
approach assuming a stationary state similar to Ref.~\onlinecite{WET00},  
which is justified if the dielectric relaxation time is short  
compared to other dynamical features.  
This provides us with the one-dimensional continuity equation 
\begin{equation} 
-\pabl{}{z}\left[ \mu n(z)\pabl{}{z}E_{F}(z)\right] 
=ef[n(z_{\mathrm{QD}}), n_{\mathrm{QD}}^{\mathrm{2D}}(t)] 
\chi_{\mathrm{QD}}(z) 
\label{EqCont} 
\end{equation} 
with the characteristic function 
\begin{eqnarray} 
   \chi_{\mathrm{QD}} ( z ) := \left\{ 
   \begin{array}{c} 
   1/h \quad (z \in \text{QD layer}) \\ 
   0 \quad (z \not\in \text{QD layer}) 
   \end{array} 
   \right. 
\end{eqnarray} 
and the electron capture rate $f$ which determines the 
electron density in the QD layer  
(in units $\mathrm{sec}^{-1} \mathrm{cm}^{-2}$) 
\begin{equation} 
\pabl{}{t}n_{\mathrm{QD}}^{\mathrm{2D}}(t)= 
f[n(z_{\mathrm{QD}}), n_{\mathrm{QD}}^{\mathrm{2D}}(t)]\, .
\end{equation} 
Here, $e>0$ is the elementary charge,  
$\mu=2 \times 10^5 \ \mathrm{cm}^2 /\mathrm{Vs}$  
is the electron mobility (at 77 K) and  
$n_{\mathrm{QD}}^{\mathrm{2D}}$ 
is the density per unit area of the electrons trapped within the QDs 
at position  
$z_{QD}$,  
$E_{F}$ is the Quasi-Fermi 
level of the electrons, which is related to the free electron density $n$ by 
\begin{equation} 
\begin{split} 
&n(z)=N_c(z) F_{1/2} \left(\frac{E_{F}(z)-E_{c0}(z)+e\phi(z)}{k_BT}\right)\\ 
&\mathrm{with}\quad 
N_c=2\left(\frac{m_e k_B T}{2 \pi \hbar^2}\right)^{3/2} 
\end{split} 
\end{equation} 
and the Fermi integral $F_{1/2}$. Here $m_e$ is the effective mass, 
$E_{c0}$ the intrinsic band edge of the conduction band 
(we set $E_{c0}=0$ for GaAs and use $E_{c0}=0.198$ meV for the 
Al$_{0.25}$Ga$_{0.75}$As layer), $T$ the 
lattice temperature,  
$k_B$ Boltzmann's constant, and $\phi$  denotes the electric potential.  
Quantization effects in the two dimensional electron gas 
are neglected for simplicity. 
 
The electron capture rate $f$  
is assumed to be a combination of Auger processes 
and single electron processes 
(like multi-phonon capture and emission) 
in the following way: 
\begin{equation} 
f[n(z_{\textrm{QD}}),n_{\mathrm{QD}}^{\mathrm{2D}}] = 
(C n^{\mathrm{2D}}+\sigma) 
(n^{\mathrm{2D}} p_{\mathrm{QD}}^{\mathrm{2D}}-n_{1}^{\mathrm{2D}} n_{\mathrm{QD}}^{\mathrm{2D}}) 
\label{EqGRraten} 
\end{equation} 
Here $n^{\mathrm{2D}}=n(z_{\textrm{QD}}) h$ is the effective free electron density 
at the QDs per unit area (which may be related to a wetting layer density 
of the same order of magnitude),  
$p_{\mathrm{QD}}^{\mathrm{2D}}=2N_{\mathrm{QD}}-n_{\mathrm{QD}}^{\mathrm{2D}}$ 
is the density of unoccupied QD states for twofold degeneracy. 
$C$ and $\sigma$ 
are the rate coefficients for the Auger and  
multi-phonon process, respectively (see also Ref.~\onlinecite{RAY00}). 
For the Auger process, electron capture requires the presence of
two electrons in the vicinity of the QD, as well as an empty
dot state. Thus the capture rate is proportional to 
$(n^{\mathrm{2D}})^2p_{\mathrm{QD}}^{\mathrm{2D}}$. In contrast, for
single electron processes the energy is transfered to different degrees
of freedom (such as phonons) and the rate is proportional to
$n^{\mathrm{2D}}p_{\mathrm{QD}}^{\mathrm{2D}}$. The respective
electron emission rates (with negative sign) are obtained from the
principle of detailed balance for equilibrium distributions \cite{SCH87}.
Assuming nondegeneracy in the conduction band, we find
$n^{\mathrm{2D}}= N_c h \exp[(E_F+e\phi)/k_BT]$ and
$n_{\mathrm{QD}}^{\mathrm{2D}}/p_{\mathrm{QD}}^{\mathrm{2D}}
=\exp[(E_F+E_b+e\phi)/k_BT]$ in equilibrium. Thus
$n^{\mathrm{2D}}p_{\mathrm{QD}}^{\mathrm{2D}}/n_{\mathrm{QD}}^{\mathrm{2D}}
=N_c h \exp(-E_b/k_BT)\equiv n_{1}^{\mathrm{2D}}$,
and the prefactor $n_{1}^{\mathrm{2D}}$ in Eq.~(\ref{EqGRraten}) 
ensures the vanishing of the rates in thermodynamic equilibrium.

The continuity equation (\ref{EqCont}) 
has to be combined with the one-dimensional Poisson equation 
\begin{equation} 
\begin{split} 
&-\epsilon_0 \pabl{}{z} \left[ \epsilon_r (z) \pabl{}{z}\phi(z) \right] = \rho ( z )\\ 
\mathrm{with}\quad 
&\rho ( z ) = e [N_D^+ ( z ) - n(z) - n_{\mathrm{QD}}^{2\mathrm{d}} (t)  
\chi_{\mathrm{QD}} ( z ) ] 
\end{split} 
\end{equation} 
where $N_D^+$ is the density of ionized donors  
resulting from the $\delta$-doping and  $\epsilon_0$ and 
$\epsilon_r$ are the absolute and relative permittivity.

The boundary condition for the quasi-Fermi level is $E_F=0$  at the 
AlGaAs/GaAs interface due the contact with the  two dimensional electron gas 
(2DEG) and 
$E_F=-eV_g$ at the GaAs/Al interface. Here the Schottky barrier 
height $E_S=1.0$ eV determines the difference between the 
conduction band edge and the Fermi level in the metal. Thus
$- e \phi (z_{\textrm{Schottky}})=E_S-eV_g$ provides 
a Dirichlet boundary condition for $\phi$. The second boundary condition 
of Neumann type results from the charge of the ionized donors 
at the $\delta$-doping. 
All calculations are performed at 77 K. 
 
\section{Results} 
 
\begin{figure}
\includegraphics[width=0.8\columnwidth]{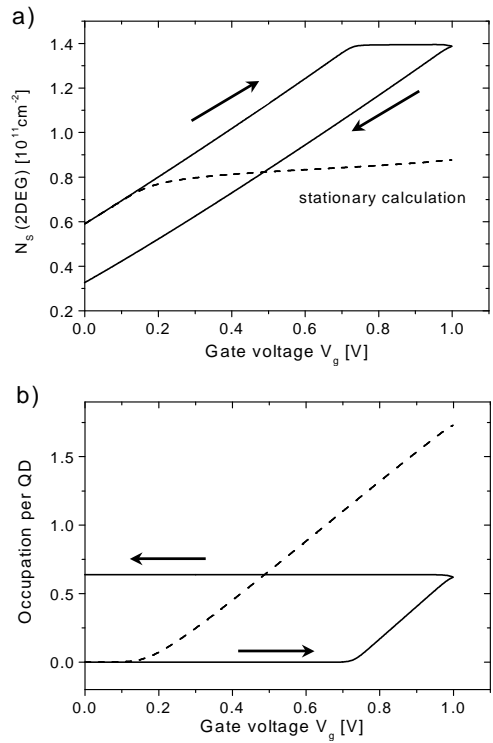} 
\caption[a]{(a) Channel electron density $N_s$ versus bias $V_g$ applied 
to the gate contact for a bias sweep. 
The duration of the entire sweep is one hour and an  Auger coefficient  
$C=2\times 10^{-12}\ \mathrm{cm}^4/s$ is used. (b) 
corresponding occupation probability of the QDs.} 
\label{FigKennAuger} 
\end{figure} 
 
At first we consider only Auger processes. We use  
$C=2\times 10^{-12}\ \mathrm{cm}^4/s$, which was calculated in 
Ref.~\onlinecite{USK98}. Fig.~\ref{FigKennAuger}(a)  shows the  density  
$N_s$ of the 2DEG versus the gate voltage 
for a bias sweep from $V_g=0$ V to 1 V and back to 0 V 
(indicated by the arrows) with a constant sweep rate and a total sweep 
duration of 1 hour \cite{Yusa02}.
The findings are in excellent 
agreement with the corresponding experimental data, see the broken line 
of Fig. 1 in Ref.~\onlinecite{YUS97}, 
except for an overall shift of the densities by 
$\sim 2\times 10^{10}/\textrm{cm}^2$ (larger density in the experiment). 
Fig.~\ref{FigKennAuger}(b) shows the corresponding occupation
probability of the QDs, demonstrating that the charging state 
of the QDs discriminates between the different conducting states. 
For a comparison we have also shown the result for the 
stationary situation $f[n,n_{\mathrm{QD}}^{\mathrm{2D}}]= 0$. 
This shows that both during  
the up- and down-sweep nonequilibrium distributions persist, as the 
GR rates are not fast enough to establish thermodynamic equilibrium. 

\begin{figure}  
\includegraphics[width=0.8\columnwidth]{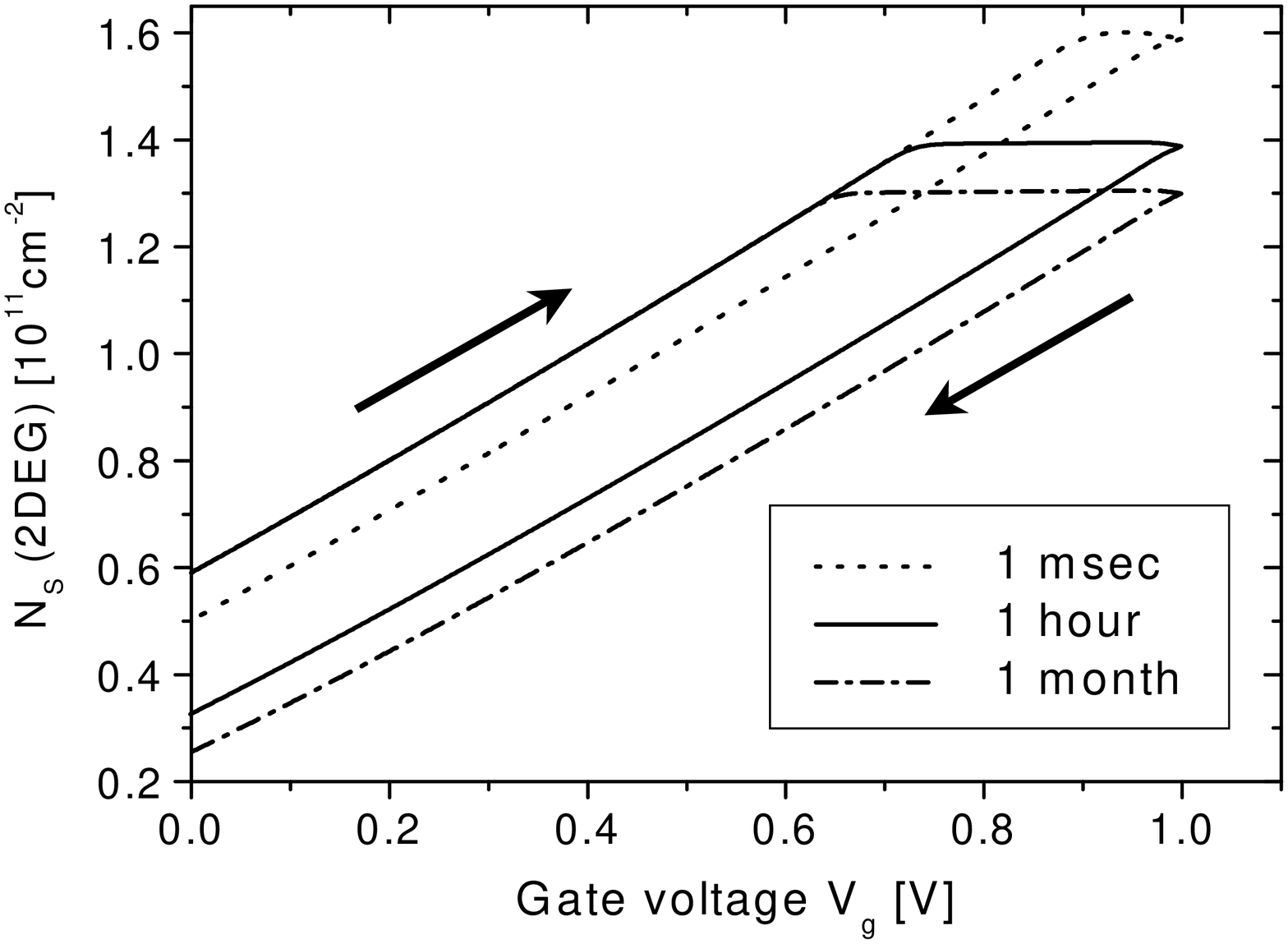} 
\caption{$N_s$-$V_g$-characteristic calculated  
for different sweep durations. An Auger coefficient 
$C=2\times 10^{-12}\ \mathrm{cm}^4/s$ is used.} 
\label{FigDiffSweeps} 
\end{figure}

The dynamical nature of the bistability manifests itself in the 
time dependence of the bias sweep, see Fig.~\ref{FigDiffSweeps}. 
This shows that the behavior is robust against changes in the time 
scales by several orders of magnitude albeit the plateau 
density where the dots are charged is slightly changing with the 
sweep duration. Below this density the charge on the QDs is 
hardly changed even for the slowest sweep rate. 
This result indicates that the branches are stable on 
rather long time scales, an important feature for possible 
applications. Time scales of 100~h have been demonstrated in  
Ref.~\cite{KOI00} at room temperature 
where a more complicated structure was used. 
 
The rather long relaxation times are due to the fact that 
Auger rates depend quadratically on the electron density $n$, see 
Eq.~(\ref{EqGRraten}).  Indeed the potential profile provides rather low 
electron densities at the QDs for biases below  
$\approx 0.7 V$, leading to a very 
small capture rate. 

The $N_s(V_g)$ characteristic can be understood as follows:
Starting from equilibrium at $V_g=0$, the quantum dots
are essentially unoccupied. This implies a linear potential
distribution over the entire GaAs layer with a potential 
$\phi(z_{\textrm{QD}})\approx (V_g-E_S/e)W/L$ at the QDs. 
Here $L$ is the distance between the 2DEG and the Schottky contact and
$W$ is the distance between the 2DEG and the QDs.
(We used  $e\phi(z_{\mathrm{2DEG}})\approx 0$ as the
Fermi level $E_F=0$ is close to the band edge at the 2DEG.)
Assuming that the electrons are in equilibrium with the 2DEG,
we have 
\begin{equation}
n(z_{\textrm{QD}})\approx N_c \exp\left(-\frac{E_S-eV_g}{k_BT}\frac{W}{L}\right)\, .
\end{equation}
For biases below $V_{\textrm{crit}}$ this density is so small
that the corresponding capture rates hardly change the
occupation of the QDs on the time scale of the bias sweep.
Therefore the linear potential drop in the GaAs layer
persists and the density of the
2DEG increases with $V_g$. 
At $V_g=V_{\textrm{crit}}$ the capture rate
becomes
\begin{equation}
f[n(z_{\textrm{QD}}),0]=\pabl{}{t}n_{\mathrm{QD}}^{\mathrm{2D}}(t)= 
\frac{\epsilon_r\epsilon_0}{e (L-W)}\frac{\d V_g}{\d t}\, .
\label{EqLevelCond}
\end{equation}
Then the charge captured in the quantum dots just screens
any further increase of the bias, so that the field on the
left side of the dots (as well as the density of the
2DEG) remains constant.
Sweeping down the bias, the electron density $n(z_{\textrm{QD}})$
decreases and so do the capture rates.
Therefore the charge in the QDs remains almost constant during
the down sweep, explaining the linear decrease of the 2DEG
electron density.

\begin{figure}
\includegraphics[width=0.8\columnwidth]{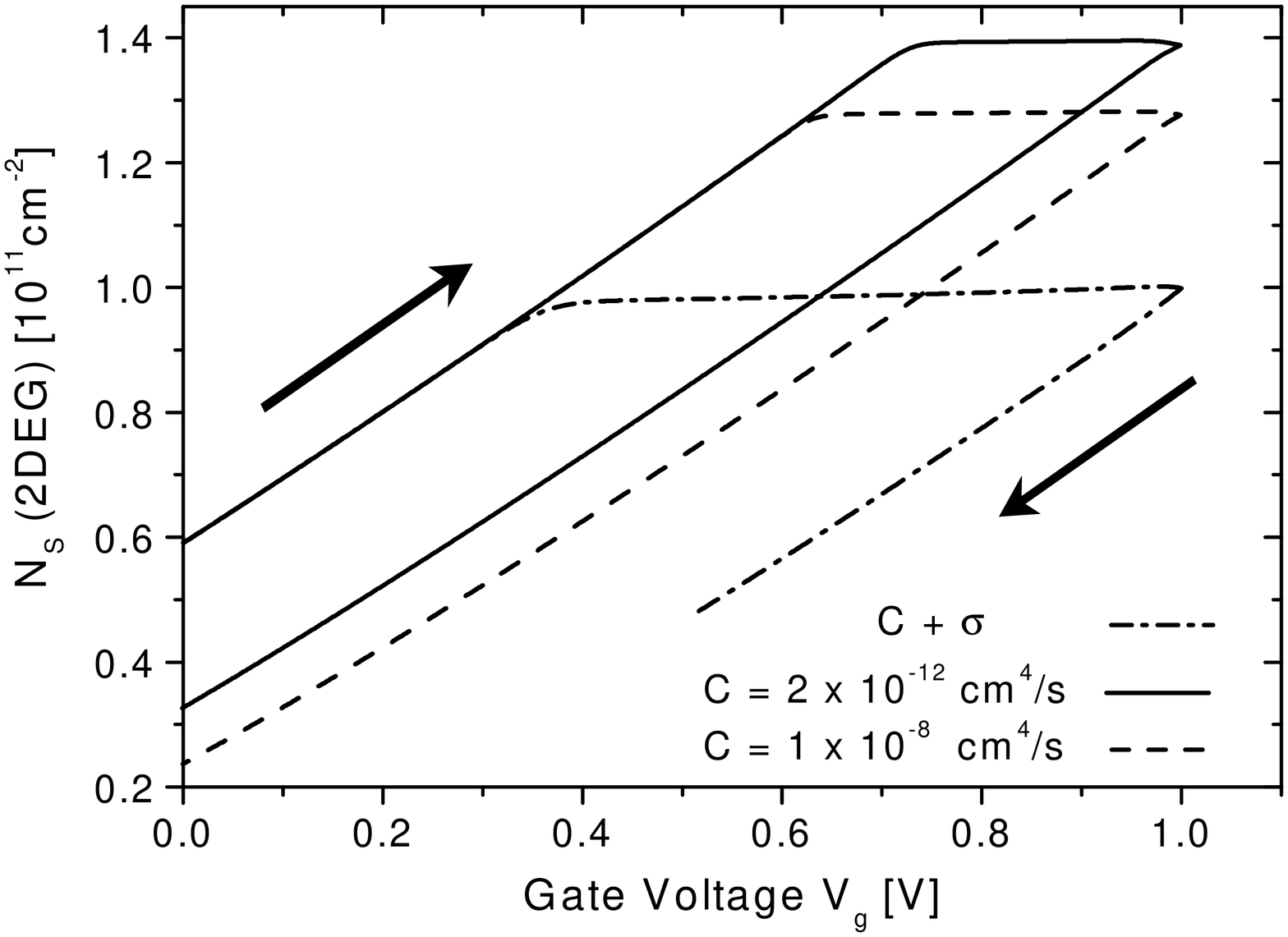} 
\caption{$N_s$-$V_g$-characteristic calculated with different  
values of the coefficient $C$ with $\sigma=0$ (full and dashed line)
as well as $C=10^{-8}\textrm{cm}^4/s$ and 
$\sigma=1\textrm{cm}^2/s$ (dash-dotted line).
The entire sweep duration is one hour.}
\label{FigDiffCoeff} 
\end{figure}

The magnitude of the  
Auger coefficient is controversial. E.g., in Ref.~\onlinecite{RAY00} 
a value of $C=10^{-8}\mathrm{cm}^4/s$  
was deduced from experimental data. 
The corresponding $N_s(V_g)$ curve is given in Fig.~\ref{FigDiffCoeff}  
(dashed curve). The result differs only slightly but agrees less well  
with the experimental data. This indicates that the findings are not 
very sensitive to the magnitude of $C$ which just rescales the time scales.
[This follows directly from Eq.~(\ref{EqLevelCond}): Only the ratio
between the capture coefficients and $\d V_g/\d t$ matters.]   
The analysis of Ref.~\onlinecite{RAY00} provides additionally a 
single electron capture coefficient $\sigma= 1\, \mathrm{cm}^2/\mathrm{s}$ 
in Eq.~(\ref{EqGRraten}). The corresponding result 
is shown by the dash-dotted line in Fig.~\ref{FigDiffCoeff}. 
The presence of the single electron term strongly changes the behavior. 
Regarding the large plateau width and its low electron density, 
this result is not compatible with the 
experiment \cite{YUS97}. 

\begin{figure}[h]  
\includegraphics[width=0.8\columnwidth]{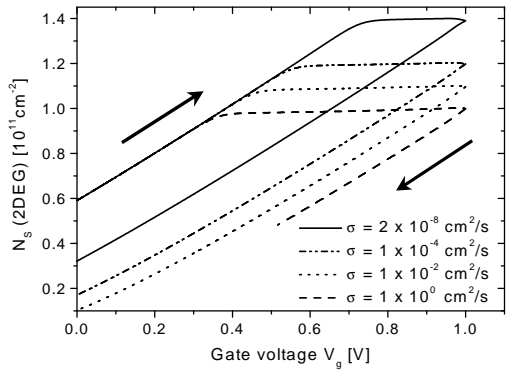} 
\caption{$N_s$-$V_g$-characteristic calculated with different values for the 
coefficient $\sigma$ and $C=0$. The duration of the entire sweep is one hour.}
\label{FigDiffCoeffPhonon} 
\end{figure}

In Fig.~(\ref{FigDiffCoeffPhonon}) the result is given for different 
single electron capture coefficients $\sigma$ while Auger processes are 
neglected (i.e. $C=0$). Good agreement with the 
experiment is found for 
$\sigma = 2 \times 10^{-8}\mathrm{cm}^2/\mathrm{s}$  
while values above $\sigma\sim 10^{-4}\mathrm{cm}^2/\mathrm{s}$ 
are in strong disagreement with the experimental findings. 
  
Experimentally, electron capture times of the order of 
$10-100$ ps for high carrier densities have been observed, 
see, e.g. Ref.~\cite{FEL01} and references cited therein. For dot densities 
of $10^{10}-10^{11}/\mathrm{cm}^2$ this implies  
$\sigma \gg 10^{-2}\mathrm{cm}^2/\mathrm{s}$ 
if single electron processes are dominating. From our study of the 
dynamical bistability this seems to be too large. 
On the other hand the Auger coefficient  
$C=2\times 10^{-12}\mathrm{cm}^4/\mathrm{s}$ 
gives capture times of 25 ps for a (wetting) layer density 
of $10^{11}/\mathrm{cm^2}$ and a dot density of 
$10^{11}/\mathrm{cm^2}$. The times become even shorter for larger 
values of $C<10^{-8}\mathrm{cm}^4/\mathrm{s}$, which are also  
consistent with our study. 
  
\section{Conclusion} 
 
We have shown that the bistability observed in quantum dot structures 
is of dynamical nature with time scales of the order of 
months under appropriate conditions. The long time scales result from the 
dependence of the  Auger capture and emission rate 
on the conduction band electron densities. For not too large gate biases, 
the densities are low and thus the kinetics is extremely slow. 
The analysis of this kind of experiments allows for 
additional information to study the relevance of Auger capture 
in comparison to single electron processes  
as multi-phonon emission. 
Our findings suggest an upper bound for single electron processes  
$\sigma < 10^{-4}\mathrm{cm}^2/\mathrm{s}$. Such values are 
not compatible with observed capture rates at high electron densities. 
In contrast,
Auger capture coefficients in the range of 
$10^{-12}\mathrm{cm}^4/\mathrm{s}<C<10^{-8}\mathrm{cm}^4/\mathrm{s}$ 
agree well with the fast dynamics for high carrier densities and  explain the 
slow dynamics for low carrier densities where the dynamical bistability is 
observed. 
 
This work was supported by DFG in the framework of SFB 296.


\end{document}